\documentclass[11pt]{article} 
\usepackage{amsmath} 
\usepackage{amsthm} 
\usepackage{amssymb}	
 \usepackage{indentfirst}
\usepackage{float}
\usepackage{titlesec}
\titlelabel{\thetitle. \quad}

\usepackage{authblk}
\usepackage{graphicx} 
\usepackage[dvips,letterpaper,margin=0.75in,bottom=0.5in]{geometry}
\usepackage[titletoc,toc,title]{appendix}
\usepackage{verbatim}
\usepackage{cite}

\newcommand{\matrixel}[3]{\left< #1 \vphantom{#2#3} \right|
 #2 \left| #3 \vphantom{#1#2} \right>} 

\begin{document}
\title{Transverse Target  Azimuthal Single Spin Asymmetry in Elastic $e-N^\uparrow$ Scattering}

\author{Tareq Alhalholy and Matthias Burkardt}
\affil{Department of Physics, New Mexico State University,
Las Cruces, NM 88003-0001, U.S.A.}
\date{\today}
\maketitle

\begin{abstract}

Using the Coulomb Eikonal Phase associated with a transversely  polarized nucleon, we calculate the azimuthal single spin asymmetry of unpolarized electrons elastically scattered from a transversely polarized nucleon target. The azimuthal asymmetry in this case is attributed to the fact that in a transversely polarized nucleon, the transverse charge density is no longer axially symmetric and is distorted due to the transverse spin of the nucleon. This asymmetry in the nucleon charge density causes a left-right  asymmetry in the azimuthal distribution of the scattered  electrons.  In our approach, we utilize the relativistic Eikonal approximation to calculate the one and two photon exchange amplitudes from which a non zero azimuthal asymmetry appears due to the interference between the two amplitudes.
\end{abstract}

\titlepage



\section{Introduction}
Polarized scattering experiments have become a primary tool to extract information about different properties of hadrons 
 in elastic and deep inelastic (DIS) scattering processes where  single (and double) spin asymmetry is the main measured quantity in the different  scattering processes. Due to time-reversal invariance, this quantity vanishes in the one photon approximation and  knowledge of the two photon exchange process is necessary to   calculate the asymmetry, beside its importance to achieve the necessary precision needed to understand discrepancies in recent experimental results related to nucleon form factors~\cite{Arrington-2PE,Carlson-TPE,Afanasev-TPE,Schlegel-TPE}.  In inclusive and semi inclusive DIS  processes, the information obtained from the observed  a zimuthal (left right) asymmetry  has attracted many current experimental and theoretical attention. Similarly in elastic $e-p$ scattering, transverse  beam or target azimuthal asymmetries  have recently become under experimental and theoretical investigation~\cite{dutta,Maas,Androic,Armstrong}, where information about nucleon form factors and nucleon structure can be extracted beside accessing the two photon exchange amplitude.\vspace{3mm} 

\noindent In this work we are interested in the azimuthal transverse (or normal) target SSA in elastic  $e-N^{\uparrow}$ scattering. A main motivation for this work is the fact that the charge density for a transversely polarized nucleon  is no longer axilay symmetric in the transverse plane and is disturbed due to the transverse spin of the nucleon~\cite{BurkardtGPDs,diehl2005}. This azimuthal asymmetry in the charge density plays a main rule in the observed asymmetry in inclusive and semi inclusive DIS~\cite{Tareq1,Burkardt2014,Metz1}. In a similar way, in elastic scattering of electrons from a transversely polarized nucleon, the azimuthal asymmetry in the nucleon charge density is expected to cause an asymmetry in the distribution of the scattered electrons.  Moreover, while  transverse target SSA in elastic scattering has been extensively  studied~\cite{Rujula,Afanasev1,TSSA-Resonance},  there is no explicit calculation of the azimuthal asymmetry in this case. In our approach, the azimuthal asymmetry of the scattered electrons appears as a  consequence of the distortion of the transverse charge density  for a transversely polarized target.\vspace{3mm}
 
\noindent In Ref.~\cite{Tareq1} the electromagnetic potetial associated  with a transversely polarized nucleon was calculated, this potential represents the impact parameter  Coulomb-Eikonal phase which carries the  asymmetry of the transverse charge density of the nucelon and represents the main input for  the  relativistic Eikonal approximation amplitude~\cite{Eik-Ads,Pol-Strass,Kabat}. By expanding the Eikonal amplitude, one can calculate the one and two photon exchange amplitudes, from which it can be shown  that SSA is indeed vanishes at the one photon exchange level, while a non zero asymmetry appears due to the interference of the one and two photon exchange amplitudes.

\section{Nucleon Transverse Charge and Magnetization Densities}
We start by a introducing the transverse structure of the nucleon~\cite{BurkardtGPDs,CarlsonVanderhaeghen1}, which is the frame of this work. For a nucleon of mass $M$ transversely polarized  (with respect to the plane of the incident  beam) in the $x$ direction, the charge density is no longer axially symmetric and the  distribution of partons in the transverse plane is  distorted which is represented by a dipole term in the parton distribution function for a transversely polarized nucleon. In terms of the relevant generalized parton distributions, GPDs, the parton distribution function is given by~\cite{BurkardtGPDs}
\begin{equation}
\begin{aligned}
q(x,\mathbf{b_\perp})  & =  \int \frac{d^2  q_\perp}{(2 \pi)^2} e^{-i \mathbf{b_\perp} \cdot \mathbf{ q_\perp}} \left[ H^q (x,0,-q^2_\perp) + i \frac{q_y}{2 M} E^q (x,0,-q^2_\perp) \right] \\
 & = q(x, b_\perp) - \frac{1}{2 M} \frac{\partial}{\partial b_y} \mathcal{E} ^q (x,\mathbf{b_\perp}),
\label{eq:PartonDist}
\end{aligned}
\end{equation}
where
\begin{equation}
q(x,b_\perp) = \int \frac{d^2 q_\perp}{(2 \pi)^2} H^q (x,0, -q^2_\perp) e^{-i \mathbf{b_\perp} \cdot \mathbf{q_ \perp}},
\end{equation}
\begin{equation}
\mathcal{E} ^q (x,b_\perp) = \int \frac{d^2 q_\perp}{(2 \pi)^2} E^q (x,-q^2_\perp) e^{-i \mathbf{b_\perp} \cdot \mathbf{q_ \perp}},
\end{equation}
using
\begin{equation}
\begin{aligned}
F_1(t) & = \sum_q e_q \int_{-1}^{1} dx H_q(x,\xi,t) , \ \  \
F_2(t) & = \sum_q e_q \int_{-1}^{1} dx E_q(x,\xi,t) , 
\label{eq:FF-GPD}
\end{aligned}
\end{equation}
and
\begin{equation}
\begin{aligned}
 \rho_1 \left( \left| \mathbf{b_\perp} \right| \right) = \int \frac{d^2 q_\perp}{\left(2 \pi \right)^2} e^{- i \mathbf{q_\perp} \cdot \mathbf{b_\perp}} F_1 \left( q^2_\perp \right), \ \
  \rho_2 \left( \left| \mathbf{b_\perp} \right| \right) = \int \frac{d^2 q_\perp}{\left(2 \pi \right)^2} e^{- i \mathbf{q_\perp} \cdot \mathbf{b_\perp}} F_2 \left( q^2_\perp \right),
  \label{eq:ChDens}
  \end{aligned}
\end{equation}
 the nucleon transverse charge density in impact parameter space can  be written as
\begin{equation}
\rho \left(\mathbf{b_\perp}\right) = \rho_1 \left( \left| \mathbf{b_\perp}\right| \right) -  \frac{1}{2 M} \frac{\partial}{\partial y} \rho_2 \left( \left| \mathbf{b_\perp}\right| \right),
\end{equation}
the second term in the above equation represents a shift in the transverse charge density in the $y$ direction with respect to the unpolarized charge density, represented by the undestorted first term in the above equation. The charge density in the transverse plane for a transversely polarized nucleon can be written in a more useful form by evaluating the Fourier transform in Eq.\eqref{eq:PartonDist}~\cite{CarlsonVanderhaeghen1}
\begin{equation}
\rho_T (\mathbf{b_\perp}) = \rho(b_\perp) - \sin(\phi_b - \phi_s) \int_{0}^{\infty} \frac{dq_\perp}{2 \pi} \frac{q^2_{\perp}}{2 M} j_1 (b_\perp q_\perp) F_2 (q^2_{\perp}).
\label{eq:Tr-Ch-Den}
\end{equation}
The above form show the explicit azimuthal dependence of the nucleon's charge density.

\section{ Coulomb Eikonal Phase and One and Two Photon Exchange Eikonal Amplitudes}
For unpolarized beam of electrons  elastically scattered  from a transversly polarized nucleon traget, the scattering amplitude in the relativistic Eikonal approximation is given by~\cite{Eik-Ads,Kabat}
\begin{equation}
f^{\uparrow}(\mathbf{q_\perp})= \ - 2 i s \int d^2b_\perp \ e^{-i\mathbf{q_\perp \cdot b_\perp}} \ \left[ e^{i\chi^{\uparrow}(\mathbf{b\perp})}-\ 1\right] ,
\label{eq:Eik1}
\end{equation}
where $s$ is the center of mass energy,  $ {\bf b_\perp} = b_\perp \cos (\phi_{b_\perp}) \ \hat{e}_x + b_\perp \sin(\phi_{b_\perp})  \ \hat{e}_y $,\ ${\bf q_\perp} = q_\perp \cos (\phi_{q_\perp}) \ \hat{e}_x + q_\perp \sin(\phi_{q_\perp}) \ \hat{e}_y$ and $\chi^{\uparrow}\left(\mathbf{{b}_\perp}\right)$ is the Coulomb/Eikonal phase associated with a nucleon target of spin transverse to the plane of the  incident beam  with spin vector $\mathbf{S} = \cos (\phi_s) \ \hat{e}_x + \sin(\phi_s) \ \hat{e}_y$.
The Coulomb/Eikonal phase is given by 
\begin{equation}
\chi^{\uparrow}(\mathbf{{b}_\perp})=\frac{- 4 \pi \alpha_e}{2 s } \ \int_{-\infty}^{\infty} dz \ A^{(0)\uparrow}(\mathbf{b_\perp},z) = \frac{- 4 \pi \alpha_e}{2 s } \ A^{(0)\uparrow}(\mathbf{b_\perp}),
\label{eq:CoulPha}
\end{equation}
where, $A^{(0)\uparrow}$ is the electromagnetic transverse potential produced by a transversely polarized nucleon of spin up, and $\alpha_e =\frac{1}{137} $ is the electromagnetic coupling constant. From Ref.~\cite{Tareq1}, the transverse potential is given by
\begin{multline}
A^{(0) \uparrow} \left( \mathbf{b_\perp} \right)  =   \frac{-1}{2 \pi}  \int_{0}^{\infty} \frac{dq_{\perp}}{q_\perp}  \left[ - J_0 \left(b_{\perp} q_\perp \right) +  J_0 \left(b_{0} q_\perp \right) \right] F_1(q^2_\perp)   + \frac{\sin\left( \phi_{b_\perp} -  \phi_s \right)}{4 \pi M}  \int_{0}^{\infty} dq_{\perp}  J_1 \left(b_{\perp} q_\perp \right)    F_2(q^2_\perp),
\label{eq:TranPot1}
\end{multline}
where $b_0$ is a reference point for the transverse potential which can be chosen to be  the transverse radius of the nucleon since the transverse potential is sufficiently vanishes outside the charge distribution~\cite{Tareq1} . By expanding the exponential in \eqref{eq:Eik1}, the Eikonal scattering amplitude takes the form
\begin{multline}
f^{\uparrow}(\mathbf{q_\perp}) = \ - 2 i s \int d^2b_\perp \  e^{-i \mathbf{q_\perp \cdot b_\perp}} \ \left[ e^{ \frac{- i 4 \pi \alpha}{2 s } A^{(0) \uparrow}(\mathbf{b_\perp}) }  - \ 1 \right] = \\
- 4 \pi \alpha \int d^2b_\perp  e^{-i\mathbf{q_\perp \cdot b_\perp}}  A^{(0) \uparrow}(\mathbf{b_\perp})  + \frac{i 8 \pi \alpha^2}{s} \int d^2b_\perp  e^{-i\mathbf{q_\perp \cdot b_\perp}}  \left[A^{(0) \uparrow}(\mathbf{b_\perp}) \right]^2 + \cdots.
\label{eq:1-2-pho-exch}
\end{multline}
\noindent The first term in the above expansion represents the one photon exchange amplitude while the second corresponds to the two photon exchange amplitude. In the following section we will explicitly calculate these two amplitudes from which one can evaluate the single spin asymmetry.

\section{Evaluation of the One and Two  Photon Exchange Eikonal Amplitudes}
The potential in Eq.\eqref{eq:TranPot1} can be written as a sum of  unpolarized and polarized terms \ $A^{(0) \uparrow}(\mathbf{b_\perp})=A^{(0) \uparrow}_u(\mathbf{b_\perp}) + A^{(0) \uparrow}_p(\mathbf{b_\perp})$, where
\begin{equation}
\begin{aligned}
A^{(0) \uparrow}_u(\mathbf{b_\perp})  & = \frac{-1}{2 \pi}  \int_{0}^{\infty} \frac{dq_{\perp}}{q_\perp}  \left[ - J_0 \left(b_{\perp} q_\perp \right) +  J_0 \left(b_{0} q_\perp \right) \right] F_1(q^2_\perp)  \\
A^{(0) \uparrow}_p(\mathbf{b_\perp})  &=\frac{\sin\left( \phi_{b_\perp} -  \phi_s \right)}{4 \pi M}  \int_{0}^{\infty} dq_{\perp}  J_1 \left(b_{\perp} q_\perp \right)    F_2(q^2_\perp).
\end{aligned}
\end{equation}
Thus, the  $1 \gamma$ amplitude (the first term in eq.\eqref{eq:1-2-pho-exch}) becomes ,
\begin{equation}
f^{\uparrow}_{(1 \gamma)}  = f^{\uparrow}_{(1 \gamma) u} \ + \ f^{\uparrow}_{(1 \gamma) p} ,
\end{equation}
where
\begin{equation}
\begin{aligned}
f^{\uparrow}_{(1 \gamma) u}(\mathbf{q_\perp})=- 4 \pi \alpha \int d^2b_\perp \ e^{-i\mathbf{q_\perp \cdot b_\perp}} \ A^{(0)\uparrow}_u(\mathbf{b_\perp}), \\
f^{\uparrow}_{(1 \gamma) p}(\mathbf{q_\perp})= - 4 \pi \alpha \int d^2b_\perp \ e^{-i\mathbf{q_\perp \cdot b_\perp}} \ A^{(0)\uparrow}_p(\mathbf{b_\perp}),
\end{aligned}
\end{equation}
the integration over \ $\phi_{b_\perp}$ \ in the unpolarized term gives  $2\pi J_0(b_\perp q_\perp)$, thus we have
\begin{equation}
\begin{aligned}
f^{\uparrow}_{(1 \gamma)u }(\mathbf{q_\perp}) & = - 8 \pi^2 \alpha  \int db_\perp \ b_\perp J_0(b_\perp q_\perp) \  A^{(0)\uparrow}_u(\mathbf{|b_\perp|}) \\
& = - 8 \pi^2 \ \alpha \  I^{(1 \gamma)}_u,
\label{eq:unpol-f1}
\end{aligned}
\end{equation}
where $ I^{(1 \gamma)}_u$ is the integral in Eq.\eqref{eq:unpol-f1}. In the polarized term,  the integration over $\phi_{b_\perp}$ \ gives\  \\ $- 2 \pi i J_1(b_\perp q_\perp) \sin{(\phi_{q_\perp}- \ \phi_{s})}$\ and the corresponding amplitude becomes
\begin{equation}
\begin{aligned}
f^{\uparrow}_{(1 \gamma) p}(\mathbf{q_\perp}) & = i 8 \pi^2 \alpha  \sin(\phi_{q_\perp} - \ \phi_{s}) \int db_\perp b_\perp J_1(b_\perp q_\perp) \ \frac{ A^{(0)\uparrow}_p(\mathbf{b_\perp})}{\sin(\phi_{b_\perp} - \ \phi_{s})}  \\ & =  i 8 \pi^2 \alpha  \sin(\phi_{q_\perp} - \ \phi_{s}) \ I^{(1 \gamma)}_p,
\label{eq:pol-f1}
\end{aligned}
\end{equation}
where $ I^{(1 \gamma)}_p$ is the integral in Eq.\eqref{eq:pol-f1}. Thus, the $1 \gamma$ amplitude becomes
\begin{equation}
\begin{aligned}
f^{\uparrow}_{(1 \gamma) }(\mathbf{q_\perp}) =  8 \pi^2 \alpha  \left[ -I^{(1 \gamma)}_u\ + \
 i  \sin(\phi_{q_\perp} - \ \phi_{s}) \ I^{(1 \gamma)}_p \right].
 \label{eq:OnePhAmp}
\end{aligned}
\end{equation}
Next we consider  the $2 \gamma$ exchange  amplitude which, from Eq.\eqref{eq:1-2-pho-exch}, is given by
\begin{equation}
f^{\uparrow}_{(2 \gamma) }(\mathbf{q_\perp})=\frac{i 8 \pi^2 \alpha^2}{s} \int d^2b_\perp \ e^{-i\mathbf{q_\perp \cdot b_\perp}} \left[ A^{(0)\uparrow} (\mathbf{b_\perp})\right]^2.
\end{equation}
Similar to the $1 \gamma$ exchange case, this amplitude can be decomposed into three parts as follows
\begin{equation}
\begin{aligned}
f^{\uparrow}_{(2 \gamma) u1}(\mathbf{q_\perp}) & = \frac{i 8 \pi^2 \alpha^2}{s} \int d^2b_\perp \ e^{-i\mathbf{q_\perp \cdot b_\perp}} \  \left[ A^{(0)\uparrow}_u (\mathbf{b_\perp}) \right]^2, \\
f^{\uparrow}_{(2 \gamma) u2}(\mathbf{q_\perp}) & = \frac{i 8 \pi^2 \alpha^2}{s} \int d^2b_\perp \ e^{-i\mathbf{q_\perp \cdot b_\perp}} \ \left[ A^{(0)\uparrow}_p (\mathbf{b_\perp}) \right]^2, \\
f^{\uparrow}_{(2 \gamma) p}(\mathbf{q_\perp}) & = \frac{i 16 \pi^2 \alpha^2}{s} \int d^2b_\perp \ e^{-i\mathbf{q_\perp \cdot b_\perp}} \  A^{(0)\uparrow}_p(\mathbf{b_\perp}) \  A^{(0)\uparrow}_u(\mathbf{b_\perp}),
\end{aligned}
\end{equation}
integrating over $\phi_{b_\perp}$ in each of the above integrals, one obtains
\begin{equation}
\begin{aligned}
f^{\uparrow}_{(2 \gamma) u1}(\mathbf{q_\perp}) & = \frac{i 16 \pi^3 \alpha^2}{s} \int db_\perp \ b_\perp J_0(b_\perp q_\perp)  \left[ A^{(0) \uparrow}_u (\mathbf{b_\perp}) \right]^2 \\
& = \frac{i 16 \pi^3 \alpha^2}{s} \ I^{(2 \gamma)}_{u1}, \\
f^{\uparrow}_{(2 \gamma) u2}(\mathbf{q_\perp}) & = \frac{i 8 \pi^3 \alpha^2}{s} \int db_\perp b_\perp \left[ J_2(b_\perp q_\perp) \cos(2 \phi_{q_\perp}-  2 \phi_{s})  +  J_0(b_\perp q_\perp) \right]  \frac{ \left[ A^{(0) \uparrow}_p (\mathbf{b_\perp}) \right]^2}{\sin^2(\phi_{b_\perp}-  \phi_{s})}\\
& =  \frac{i 8 \pi^3 \alpha^2}{s} \left[ I^{(2 \gamma)}_{u_{2a}} \ \cos(2 \phi_{q_\perp} - \ 2 \phi_{s}) \ + \  I^{(2 \gamma)}_{u_{2b}} \  \right], \\
f^{\uparrow}_{(2 \gamma) p}(\mathbf{q_\perp}) & = \frac{ 32 \pi^3 \alpha^2}{s} \sin(\phi_{q_\perp} - \ \phi_{s}) \int db_\perp  \ b_\perp J_1(b_\perp q_\perp) \ \frac{A^{(0) \uparrow}_p (\mathbf{b_\perp})}{\sin(\phi_{b_\perp} - \ \phi_{s})} \   A^{(0) \uparrow}_u (\mathbf{b_\perp})\\
& = \frac{ 32 \pi^3 \alpha^2}{s} \ \sin(\phi_{q_\perp} - \ \phi_{s}) \ I^{(2 \gamma)}_p ,
\end{aligned}
\end{equation}
where in the second integral we  used
$$
\int_{0}^{2 \pi} d \phi_{b_\perp} \sin^2(\phi_{b_\perp} -  \phi_{s})  e^{(-iq_\perp b_\perp \cos (\phi_{b_\perp} -  \phi_{q_\perp}))}=\pi  \left[ J_2(b_\perp q_\perp) \cos (2 \phi_{q_\perp} -  2 \phi_{s})+J_0(b_\perp q_\perp) \right].
$$
Therefore, the $2\gamma$ exchange amplitude becomes
\begin{equation}
f^{\uparrow}_{(2 \gamma)}(\mathbf{q_\perp}) =  \frac{16 \pi^3 \alpha^2}{s} \left[ \ i   \left( I^{(2 \gamma)}_{u1} + \frac{I^{(2 \gamma)}_{u_{2a}}}{2} \ \cos(2 \phi_{q_\perp} - \ 2 \phi_{s}) \ + \ \frac{1}{2} I^{(2 \gamma)}_{u_{2b}} \right) +  2 \ I^{(2 \gamma)}_p \ \sin(\phi_{q_\perp} - \ \phi_{s})   \right] 
\label{eq:TwoPhAmp}
\end{equation}
From Eqs.(\ref{eq:OnePhAmp},\ref{eq:TwoPhAmp}) it is clear that the one and two photon amplitudes are azimuthally asymmetric, however the corresponding cross sections are azimuthally symmetric while the azimuthal asymmetry appears in the interference term between the $1\gamma$ and $2\gamma$ amplitudes as we shall see in the next section. 
\section{Azimuthal Transverse Target Single Spin  Asymmetry in Elastic $e - p^\uparrow$ Scattering }
Single spin asymmetry is defined as
\begin{equation}
A_n \ = \ \frac{\sigma^{\uparrow} - \sigma^{\downarrow}}{\sigma^{\uparrow} + \sigma^{\downarrow}} ,
\label{eq:SSA}
\end{equation}
The spin up and spin down cross section are given by
\begin{equation}
\begin{aligned}
\sigma^{\uparrow} & = |f^{\uparrow}_{1\gamma} + f^{\uparrow}_{2\gamma}|^2 = |f^{\uparrow}_{1\gamma}|^2 + |f^{\uparrow}_{2\gamma}|^2 + f^{* \uparrow}_{1\gamma}f^{ \uparrow}_{2\gamma} + f^{\uparrow}_{1\gamma}f^{* \uparrow}_{2\gamma}, \\
\sigma^{\downarrow} & = |f^{\downarrow}_{1\gamma} + f^{\downarrow}_{2\gamma}|^2 = |f^{\downarrow}_{1\gamma}|^2 + |f^{\downarrow}_{2\gamma}|^2 + f^{* \downarrow}_{1\gamma}f^{ \downarrow}_{2\gamma} + f^{\downarrow}_{1\gamma}f^{* \downarrow}_{2\gamma},
\label{eq:CrossSec} 
\end{aligned}
\end{equation}
 The spin down amplitude is obtained by rotating the spin vector by $\pi$. Thus, from Eqs.(\ref{eq:OnePhAmp}, \ref{eq:TwoPhAmp}), one obtains
\begin{equation}
\begin{aligned}
f^{* \uparrow}_{1\gamma} & = f^{\downarrow}_{1\gamma} \ \ \ \Rightarrow \ \ f^{* \downarrow}_{1\gamma} = f^{\uparrow}_{1\gamma} \\
f^{* \uparrow}_{2\gamma} & = -f^{\downarrow}_{2\gamma} \ \Rightarrow \ f^{ \uparrow}_{2\gamma} = -f^{* \downarrow}_{2\gamma}
\end{aligned}.
\end{equation}
Therefore, we get
\begin{equation}
\begin{aligned}
\sigma^{\uparrow} - \sigma^{\downarrow} & = 2 f^{* \uparrow}_{1\gamma} f^{\uparrow}_{2\gamma} + 2 f^{ \uparrow}_{1\gamma} f^{* \uparrow}_{2\gamma}, \\
\sigma^{\uparrow} + \sigma^{\downarrow} & = 2 |f^{\uparrow}_{1\gamma}|^2 + 2 |f^{\uparrow}_{2\gamma}|^2 ,
\end{aligned}
\end{equation} 
and the asymmetry takes the form
\begin{equation}
A_n = \frac{2 Re [f^{* \uparrow}_{1\gamma}f^{\uparrow}_{2\gamma}]}{|f^{\uparrow}_{1\gamma}|^2 +  |f^{\uparrow}_{2\gamma}|^2}.
\end{equation}
The figures below show the proton transverse target azimuthal SSA for various center of mass energies and momentum transfer. From the figures one notes that the asymmetry is positive and it increases with $q_\perp$ but decreases with the center of mass energy,as expected. On the other hand, the magnitudes of the amplitudes in the figures are consistent with the expected values for the transverse target SSA with unpolarized electrons beam~\cite{Carlson-TPE}. 

\begin{figure}[H]
 \begin{center}
  \includegraphics[width=0.45\textwidth]{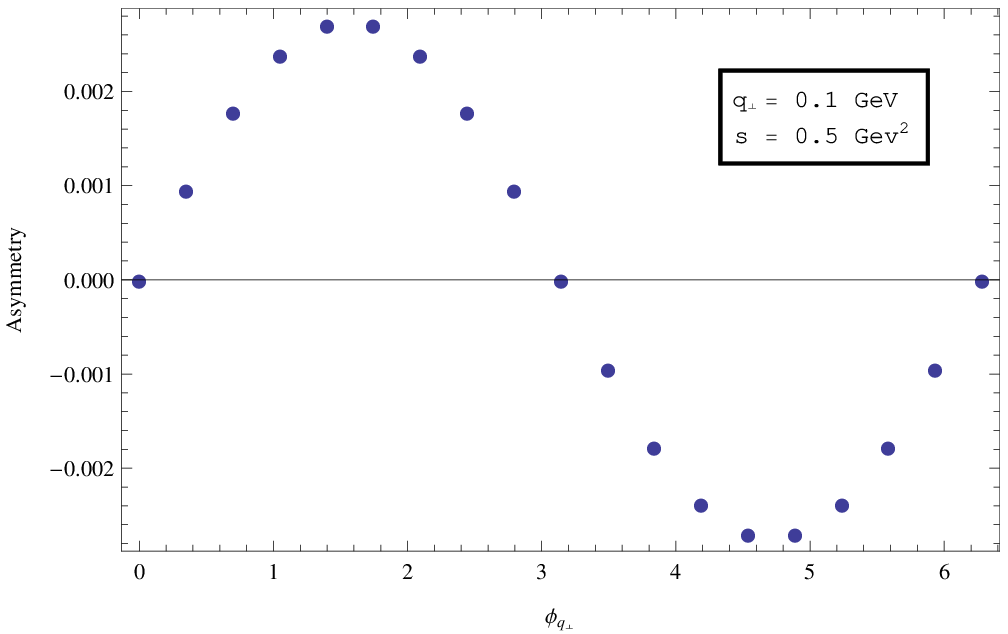}
  \includegraphics[width=0.45\textwidth]{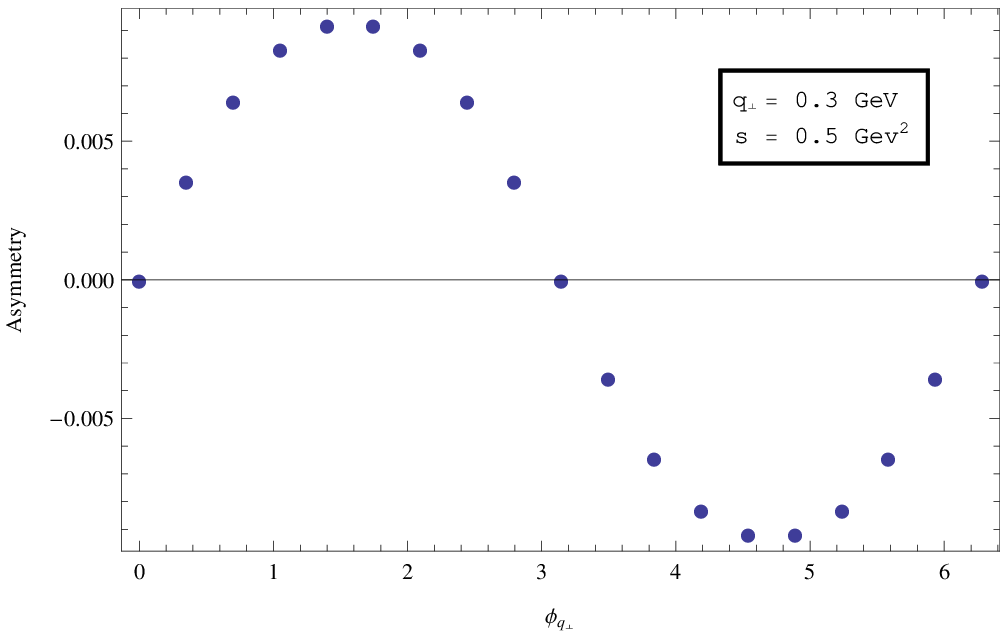}
  \includegraphics[width=0.45\textwidth]{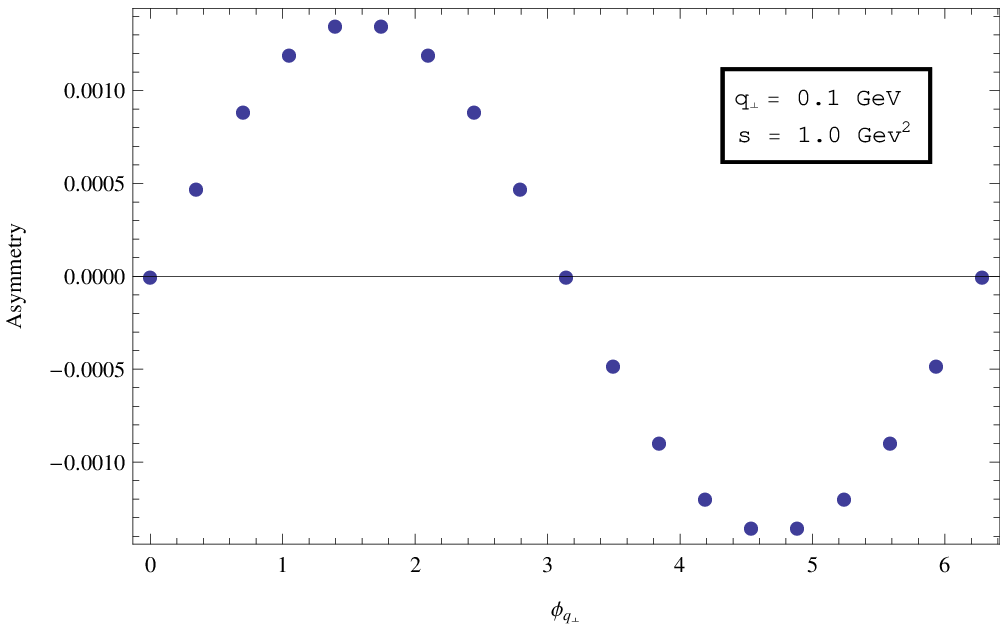}
  \includegraphics[width=0.45\textwidth]{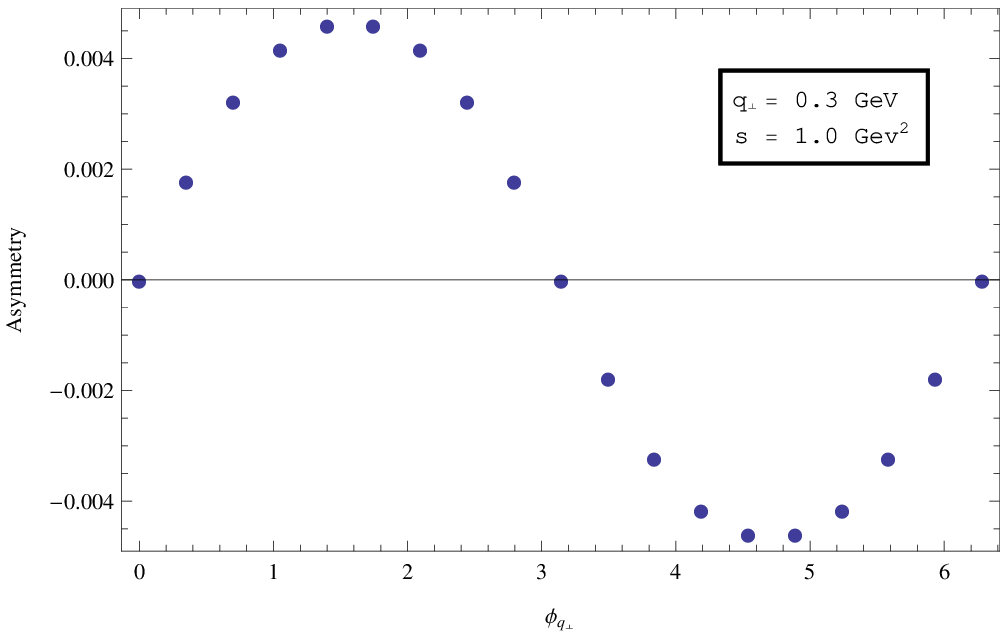}
\caption{Proton transverse target azimuthal SSA for various center of mass energies and momentum transfers using GPD parametrization for the proton form factors}
 \label{fig:TTSSA}
 \end{center}
\end{figure}

\section{Conclusion}
 We study single spin azimuthal asymmetry in elastic  electron-nucleon scattering for the case of transversely polarized nucleon with unpolarized electrons beam. The asymmetry  in this case appears due to the interference between the $1 \gamma$ and $2 \gamma$ exchange amplitudes.
To calculate the $(1)$ and $(2 \gamma)$  Eikonal scattering amplitudes, we used  the Coulomb/Eikonal phase associated with a transversely polarized nucleon, this phase is asymmetric in impact parameter space; consistent with the fact that the nucleon charge density (or the impact parameter dependent parton distribution function) is transversely  distorted due to the  transverse spin of the nucleon, which explains  the existence of  SSA  in this case.
Figure.\ref{fig:TTSSA} shows the azimuthal target single spin asymmetry for different values of the center of mass energy and momentum transfer, as expected, the asymmetry increases with $q_\perp$ and decreases with the center of mass energy. The apmlitudes in the figures are consistent with the expected values of the transverse target SSA with unpolarized electrons beam.   In the calculations, the transverse radius of the proton was taken from Ref.~\cite{CarlsonVanderhaeghen1} and the  GPD parametrization of the proton form factors from Ref.~\cite{GPD2} while the integrals were evaluated numerically using the CUBA library for multidimensional numerical integration~\cite{CUBA}.

\appendixtitleon
\appendixtitletocon

\begin{appendices}

\section{Parametrization of Form Factors Using Generalized Parton Distributions}

\vspace{3mm}

 Generalized parton distribution (GPDs) can be considered as generalization of ordinary parton distributions. The formal definition of GPDs for transversely polarized nucleon but unpolarized quarks is given by~\cite{Miller}
\begin{equation}
\begin{aligned}
\matrixel{p',S'}{\hat{O}_q(x, \textbf{b}_\perp)}{p,S}\ = \ \frac{1}{2 \bar{P}^+} \bar{u}(P',S') \left(\gamma^{+}\ H_q(x,\xi,t) \ + \ i\frac{\sigma^{+ \nu} \Delta_\nu}{2M} E_q(x,\xi,t)  \right)  u(p,S)
\end{aligned}
\end{equation}
Where $\bar{P}^\mu = \frac{1}{2}(P^\mu + P'^\mu)$ represents the average momentum of the target, $q^\mu = P'^\mu - P^\mu$ is the four momentum transfer, $t=q^2$ is the invariant momentum transfer and $\xi = - \frac{q^+}{2 P^+}$ is the change in the longitudinal component of the target momentum and is called the skewness. The nucleons form factors can be decomposed as follows 
\begin{equation}
\begin{aligned}
F^p_i = e_u F^u_i + e_d F^d_i + e_s F^s_i, \ \  \ \
F^n_i = e_u F^d_i + e_d F^u_i + e_s F^s_i
\end{aligned}
\end{equation}
Where $i = 1, 2$ and \ $e_u=\frac{2}{3}, \ e_d = e_s= \frac{-1}{3}$. The Dirac and Pauli flavor form factors at zero skewness are give by the following sum rules
\begin{equation}
\begin{aligned}
F^q_1(t) = \int_{-1}^1 dx \ H^q (x,\xi,t),  \ \  \ \
F^q_2(t) = \int_{-1}^1 dx \ E^q (x,\xi,t)
\end{aligned}
\end{equation}
\noindent The result of integration is independent of $\xi$. Also the integration region can be reduced to $0 < x < 1$ by introducing the non-forward parton densities
\begin{equation}
\begin{aligned}
\mathcal{H}^q (x,t) & = H^q (x,0,t) +  H^q (-x,0,t), \ \ \ \
\mathcal{E}^q (x,t)  = E^q (x,0,t) +  E^q (-x,0,t)
\end{aligned}
\end{equation}
Where $q= u, d$ and $\mathcal{H}^q (x,t)$ reduces to the usual valence quark densities for $t \rightarrow 0$ for the up and down quarks. Thus, the form factors become
\begin{equation}
\begin{aligned}
F^q_1(t)  =  \int_{0}^1 dx \ \mathcal{H}^q (x,t), \ \ \ \ 
F^q_2(t)  =  \int_{0}^1 dx \ \mathcal{E}^q (x,t) 
\end{aligned}
\end{equation}
The magnetic densities satisfies the following normalization conditions
\begin{equation}
\begin{aligned}
\kappa_q  =  \int_{0}^1 dx \ \mathcal{E}^q (x), \ \ \
\kappa_u & = 2 \kappa_p + \kappa_n = + 1.673, \ \ \ 
\kappa_d  = \kappa_p + 2 \kappa_n = - 2.033 \\
F^p_2   (t=0) & = 1.793, \ \ \ \ 
F^n_2  (t=0)  = - 1.913
\end{aligned}
\end{equation}
Following~\cite{GPD2}, the anzats for the $GPDs$ are
The ansatz for the $GPDs$ is~\cite{GPD2}
\begin{equation}
\begin{aligned}
\mathcal{H}^q (x,t) & = q_v (x) x^{- \alpha' (1-x) t}, \\
\mathcal{E}^q  (x,t) & = \frac{\kappa_q}{N_q} \left (1-x \right)^{\eta_q} q_v x^{- \alpha' (1-x) t},
\end{aligned}
\end{equation}
The normalization constants $N_q$ satisfies
\begin{equation}
N_q = \int _0^1 dx (1-x)^{\eta_q} q_{\nu} (x),
\end{equation}
and the  unpolarized parton distributions are parametrized as
\begin{equation}
\begin{aligned}
u_\nu (x) & = 0.262 x^{-0.69} (1-x)^{3.50} (1+3.83 x^{0.5} + 37.65 x) \\
d_\nu (x) & = 0.061 x^{-0.65} (1-x)^{4.03} (1+49.05 x^{0.5} + 8.65 x).
\end{aligned}
\end{equation}
The parameters used in this fit are $ \alpha' = 1.105, \eta_u = 1.713, \eta_d = 0.566$. The following is a plot of $F_1$ and $F_2$ as a function of the momentum transfer.

\end{appendices}

\vspace{5mm}
\bibliographystyle{unsrt}
\bibliography{TTSSA.bib}

\end{document}